\documentclass[onecolumn,floatfix,aps,nofootinbib]{revtex4}
\usepackage[dvips]{epsfig}
\usepackage[english]{babel}
\usepackage{bbm}
\usepackage{verbatim}
\usepackage{mathtools}
\usepackage{array}
\usepackage{bm} 
\usepackage{amsmath}
\usepackage{amsfonts}
\usepackage{amssymb}
\usepackage{hyperref}
\usepackage[dvipsnames]{xcolor}
\hypersetup{colorlinks=true,linkbordercolor=Blue,linkcolor=Blue, citecolor=Blue}
\usepackage{indentfirst} %primeiro paragrafo com espaço
\DeclareMathOperator{\End}{End}

\usepackage{bbold}

\begin{document}

\title{Spinorial discrete symmetries and adjoint structures}

	\author{J. M. Hoff da Silva}
	\email{julio.hoff@unesp.br}
	\affiliation{Departamento de F\'{\i}sica, \\ Universidade Estadual Paulista, UNESP,\\ Guaratinguet\'{a}, SP, Brazil.}
	
    \author{R. J. Bueno Rogerio}
	\email{rodolforogerio@gmail.com}
	\affiliation{Instituto de F\'isica e Qu\'imica, Universidade Federal de Itajub\'a - IFQ/UNIFEI, Av. BPS 1303, CEP 37500-903, Itajub\'a - MG, Brazil.}

	\author{N. C. R. Quinquiolo}
	\email{natan.quinquiolo@unesp.br}
	\affiliation{Departamento de F\'{\i}sica, \\ Universidade Estadual Paulista, UNESP, \\ Guaratinguet\'{a}, SP, Brazil.}

	\begin{abstract}
		Extending the investigations about the theory of duals, we analyze duals built up with the aid of discrete symmetry operators. We scrutinize algebraic and physical constraints (encompassing them in a theoretical scope) in order to verify which combination of discrete symmetries may compose a physical and mathematical well-posed spinorial dual. In this scenario, we relate the Lounesto classification with several other spinor classification possibilities, attempting to connect classes and physical constraints.  Several possibilities are investigated.  
	\end{abstract}

	\maketitle
	
	\section{Introduction}
	
The great success of Dirac fields describing spin 1/2 particles with fermionic statistic in the Standard Model is certainly one of the aspects which lead part of the physical community to the appreciation of spinors, Clifford algebra, and correlated issues. However, the completeness and far-reaching results of this standard formulation shall not be confused with inevitability. A closer inspection into Clifford algebras and spinors definitions \cite{lou,che} shows the existence of some particular choices, most of them suitable and consistent, but still choices, in the whole formulation whose abandonment in favor of something different can still lead to a consistent theoretical structuring, as far as physical and algebraic consequences are considered.

Despite the importance and richness carried by Dirac spinors, mathematically speaking its definition and the dual structure are presented superficially without separating necessities from conventions. However, if other spinorial physical fields describing relevant particles exist, it is necessary to know whether they have the same dual structure as the one used for Dirac spinors or not. Such an issue is commonly put aside. Undertaking a deep analysis of spinorial duals could help us to get closer to answering the above-mentioned question. In parallel, the theoretical discovery of the so-called mass dimension one fermions \cite{Ahl} draws attention because of its peculiar structure and physical impact --- being widely explored in cosmology, mathematical physics, and phenomenology framework \cite{saulo1, saulo2, saulo3, fab1, fab2, fabpolar, fabaaca, CY0, CYlag, CYM, CYST,CY21} --- making it necessary to revisit most of the fundamental aspects of the quantum field theory (QFT) and, for instance, spinorial dual theory, looking towards retrieve relevant physical information \cite{CYlag, epl}. By using a general spinor playing the role of expansion coefficient function of a spin 1/2 quantum field, the right appreciation of a specific dual structure may lead to a local quantum field within a theory respecting Lorentz symmetries, with spinors automatically restricted to a certain class, according to Lounesto \cite{lou}. Thus, the right appreciation of the dual structure, may break barriers and lead to paths still little explored in the context of QFT. This is the crevice we shall explore here. 

In particular, the above context applies to the dual spinorial theory (see \cite{Ahl} for a nice account of the case of Elko duals). In Ref. \cite{eprj} generalizations of duals were pursued and a mapping structure was settled, presenting conditions under which the operators composing new duals do form a group. Here we shall further explore this program by investigating operators entering in the spinorial dual formulation which are representations of spin 1/2 $SL(2,\mathbb{C})$ discrete symmetries $P$ and $T$ along with internal charge conjugation $C$ symmetry. We recall and use the constraints presented in \cite{eprj}, framing it in an adequate mathematical scope and extending it showing, in addiction, the algebraic constraints covariance. Going further we present a complete account of the $C,P,T$ usage in the composition of duals, presenting also possible ways to eventually circumvent some formal, but not inevitable, constraints. Within the analysis performed, the standard Dirac dual appears as the simplest dual possible. Besides, some impacts on quantum field theory, such as the relation between spinor dual theory and spin-statistic, are depicted.    

The paper is organized as follows: in the next section, a mathematical inclined general discussion is presented. In section \ref{III}, the discrete symmetry duals are studied. Section \ref{IV} is reserved to investigate some possible extensions of the usual Lounesto classification and in section \ref{remarks} we conclude with several remarks performing an outlook of consequences and questions which deserves further exploration.  
	
	\section{General Discussion}\label{II}
	
Let $\mathbb{R}^{1,3}$ be the Minkowski space and $Cl_{1,3}$ the associative unital Clifford algebra. Let also $\gamma$ be the Clifford mapping $\gamma:\mathbb{R}^{1,3}\rightarrow Cl_{1,3}$ with Clifford product given by 
\begin{eqnarray}
\gamma(u)\gamma(v)=\gamma(u)\cdot\gamma(v)+\gamma(u)\wedge\gamma(v),\label{l1}
\end{eqnarray} where $u,v$ are Minkowski space arbitrary vectors. As usual $\{\gamma(u),\gamma(v)\}=2\eta(u,v)$, where $\{\cdot,\cdot\}$ denotes the anticomutator and $\eta\in (\mathbb{R}^{1,3})^*\otimes(\mathbb{R}^{1,3})^*$ stands for the Minkowski metric. For every $\{e_\mu\}$ base of $\mathbb{R}^{1,3}$ and $\{dx^\nu\}$ base of $(\mathbb{R}^{1,3})^*$ (such that $dx^ \nu(e_\mu)=\delta^\nu_\mu$) the linearity of $\gamma$ allows one to write $\{\gamma(e_\mu),\gamma(e_\nu)\}=2\eta_{\alpha\beta}dx^\alpha\otimes dx^\beta(e_\mu,e_\nu)$, where in this paper we assume $\eta_{\mu\nu}$ as the mostly negative entries of a diagonal matrix. The standard notation $\gamma(e_\mu)\equiv \gamma_\mu$ leads then to the overspread Clifford algebra fundamental relation $\{\gamma_\mu,\gamma_\nu\}=2\eta_{\mu\nu}$. 

Let $F$ be a primitive idempotent. Algebraic spinors may be faced as proper ideals of a Clifford algebra in such a way that minimal left ideals as $Cl_{1,3}F$ encompass usual spinors, say $\Psi$, while minimal right ideals $FCl_{1,3}$ comprise dual spinors $\Psi^\star$. In this perspective, scalars are elements of $FCl_{1,3}F$ obtained by means of an inner product $\pi:FCl_{1,3}\times Cl_{1,3}F\rightarrow FCl_{1,3}F\simeq \mathbb{R}$. A slight, but important, modification is obtained employing the complexified Clifford algebra $\mathbb{C}\otimes Cl_{1,3}$. Within this context $\mathbb{C}\otimes Cl_{1,3}F\supset \Psi$, $F\mathbb{C}\otimes Cl_{1,3}\supset \Psi^*$, and $\pi:F\mathbb{C}\otimes Cl_{1,3}\times \mathbb{C}\otimes Cl_{1,3}F\rightarrow F\mathbb{C}\otimes Cl_{1,3}F\simeq \mathbb{C}$.    	

It is usually taken as a benefit for a physical theory dealing with spinors to relate left and right ideals. In fact, it would be quite an odd idea to give up on a one-to-one algebraic bridge between spinors and their duals. In such a case the construction of physical observable sets would be jeopardized. Algebraic involutions are the natural objects used to build these bridges, but they do not preserve idempotents. Therefore, in order to properly ensure a dual $\Psi^*$ for a given spinor $\Psi$ is necessary an additional ingredient. Let $\iota: \mathbb{C}\otimes Cl_{1,3}\rightarrow \mathbb{C}\otimes Cl_{1,3}$ be an involution already endowed with complex conjugation. The fact that $\iota(\mathbb{C}\otimes Cl_{1,3}F)=\iota(F)\mathbb{C}^*\otimes Cl_{1,3}$ with $\iota(F)\neq F$ in general -- the alluded no preservation of idempotents -- is fixed by means of an hermitian operator $Q \in \mathbb{C}\otimes Cl_{1,3}$ such that $\iota(a)=Q^{-1}a^\dagger Q$. In fact, note that 
\begin{eqnarray}
Q\iota(\mathbb{C}\otimes Cl_{1,3}F)=Q\iota(F)\mathbb{C}^*\otimes Cl_{1,3}=FQ(\mathbb{C}^*\otimes Cl_{1,3}). \label{l2}
\end{eqnarray} Therefore an element identified to $Q\iota(\mathbb{C}\otimes Cl_{1,3}F)$ belongs to a right ideal and comprises then a dual spinor. Hence 
\begin{equation}
\Psi^\star=Q\iota(\Psi)=QQ^{-1}\Psi^\dagger Q=(Q\Psi)^\dagger.\label{l3}
\end{equation} 

In the determination of $Q$, physical symmetry concepts are relevant. In order to appreciate them, let us face the spinor concept from the physical perspective. A relativistic spinor may be understood as a section of the $P_{SL(2,\mathbb{C})}\times_\rho\mathbb{C}^4$ bundle\footnote{The definitions of spinors used in this paper are all equivalent \cite{rol}.}, where for this paper we fix $\rho=(1/2,0)\oplus(0,1/2)$, so that the spinorial object carrying a spin $1/2$ representation of the Lorentz group belongs to the entire representation space. That is to say, the spinor is composed of left- and right-hand Weyl spinors. This last aspect is crucial to the specification of the $Q$ operator and shall not be underestimated. As expected, the scalar obtained from $\pi(\Psi,\Psi)=\Psi^*\Psi$ (the norm) shall be Lorentz invariant. Calling $Q=\eta\Delta$ we have, from (\ref{l3}), $\pi(\Psi,\Psi)=(\Delta\Psi)^\dagger\eta\Psi$. We shall approach a more general condition coming from specific considerations of $\Delta$ in the next section. By now we just take $\Delta=\mathbbm{1}$ (the identity) for a moment and remark that, as shown in \cite{Ahl}, invariance of the norm with respect to rotations and boosts does not fix completely $\eta$. Additionally, the requirements of norm invariance under parity (or time reversal, or charge conjugation) and reality set\footnote{Apart from another relevant freedom \cite{Ahl,epl,pla} which we shall not discuss here.} $\eta=\gamma_0$. The Dirac dual is then reached, and both sectors of the representation space are treated on an equal footing. 

In the next section we shall investigate $Q$ operators obtained by the appreciation of different discrete symmetries in the $\Delta$ sector, keeping $\eta=\gamma_0$, as well as combinations of discrete symmetries. While it is still possible, in general, other forms to $\eta$ along with $\Delta$ different from the identity, it is quite convenient to set $Q=\gamma^0\Delta$ since usual expressions involving spinorial transformations may be used.   
   	
\section{Algebraic and Physical constraints}\label{III}	
	
From the previous section discussion, we have arrived at $\Psi^\star=(\Delta \Psi)^\dagger \gamma^0$ by setting $Q=\gamma^0\Delta$. The hermiticity of $Q$, necessary to ensure $\iota\circ\iota(a)=a$, $\forall a\in \mathbb{C}\otimes Cl_{1,3}$, leads to the following algebraic constraint $\Delta^\dagger\gamma^0=\gamma^0\Delta$ to be respected for all $\Delta$ candidates. Let $S(\Lambda)$ denotes a spin $1/2$ Lorentz transformation so that spinors are simply transformed as $\Psi'=S(\Lambda)\Psi$. By using the standard relation $S^{-1}=\gamma^0S^\dagger\gamma^0$, it is straightforward to see that the transformed norm may be written as 
\begin{equation}      	
	\Psi'^\star\Psi'=\Psi^\dagger S^\dagger\Delta'^\dagger(S^{-1})^\dagger\gamma^0\Psi,\label{l4}
\end{equation} where $\Delta'$ stands for the transformed $\Delta$ operator. When $[\Delta,S]=0$ a sufficient condition to the norm invariance is reached. However, when this is not the case we can rely on the following statement: the spinorial norm composed by a given non-standard dual, as defined before, is invariant if, and only if, it transforms according to $\Delta'=S\Delta S^{-1}$. In fact, starting from the operator transformation it can be readily seen that $\Delta^\dagger=S^\dagger\Delta'^\dagger (S^{-1})^\dagger$. Inserting it back into (\ref{l4}) the invariance is obtained. Reciprocally, the requirement $\Psi'^\star\Psi'=\Psi^\star\Psi$ leads to
\begin{equation}
\Psi^\dagger\{S^\dagger\Delta'^\dagger(S^{-1})^\dagger-\Delta^\dagger\}\gamma^0\Psi=0.\label{l6}
\end{equation} In order to (\ref{l6}) be satisfied regardless the spinor at hands, we are forced to conclude that $\Delta^\dagger=S^\dagger\Delta'^\dagger (S^{-1})^\dagger$ and hence $\Delta'=S\Delta S^{-1}$. Note that with these results we may express the dual transformation as $\Psi'^{\star}=\Psi^\dagger\Delta^\dagger\gamma^0S^{-1}=\Psi^\star S^{-1}$, a sufficient condition to ensure the right covariance to all bilinear.  

There is an important corollary coming from the fact that $\Delta$ transformation goes as $\Delta'=S\Delta S^{-1}$. Expressing it again as $\Delta^\dagger=S^\dagger\Delta'^\dagger (S^{-1})^\dagger$ and making use of $S^{-1}=\gamma^0S^\dagger\gamma^0$ we arrive at $\Delta'^\dagger=\gamma^0S\gamma^0\Delta^\dagger S^\dagger$. By properly equating some identities we have
\begin{equation} \Delta'^\dagger=\gamma^0S(\gamma^0\Delta^\dagger\gamma^0)(\gamma^0S^\dagger\gamma^0)\gamma^0, \label{l7}
\end{equation} which may be recast as $\Delta'^\dagger=\gamma^0S\Delta S^{-1}\gamma^0$, or simply $\Delta'^\dagger=\gamma^0\Delta'\gamma^0$. But this is to say that the algebraic constraint is covariant. Indeed a relevant consistency result for the relativist output of this algebraic formulation needs to preserve. We have shown the covariance of the algebraic constraint starting from the $\Delta$ transformation law. As a further check, we may proceed to start from $\Delta^\dagger=\gamma^0\Delta\gamma^0$ and note that
\begin{equation}
S\Delta S^{-1}=S\gamma^0\Delta^\dagger\gamma^0S^{-1}.\label{l8}
\end{equation} Therefore, recognizing the left-hand side as $\Delta'$, a simple algebra leads to $\Delta'=\gamma^0(S\Delta S^{-1})^\dagger\gamma^0$, from which the covariance of the constraint is again manifest. 

Now we can investigate, from among different $\Delta$ operators within the scope of discrete spinorial symmetries, which may serve to compose consistent duals in the sense that it respects both, the algebraic and the covariance, constraints. In the following, we use Dirac matrices as given by  
\begin{equation}
\gamma_0=\begin{pmatrix}
0_{2\times2} & 1_{2\times2} \\ 1_{2\times2} & 0_{2\times2}  
\end{pmatrix},\\
\gamma_i=\begin{pmatrix}
0_{2\times2} & \sigma_i \\ -\sigma_i & 0_{2\times2}  
\end{pmatrix}, \\ \gamma_5=\begin{pmatrix}
1_{2\times2} & 0_{2\times2} \\ 0_{2\times2} & -1_{2\times2}  
\end{pmatrix},
\end{equation} where $\sigma_i$ $(i=1,2,3)$ are the standard Pauli matrices. We start from the relevant case of parity. Firstly we just recall that when acting upon spinors of arbitrary momentum, there is a correspondence between parity and Dirac operator \cite{loh}, i. e., $P \in \End(P_{SL(2,\mathbb{C})}\times_\rho\mathbb{C}^4)$ such that $P=m^{-1}\gamma^\mu p_\mu$. Of course, in the momentum space $p^\mu$ is a simple coordinate, and therefore $\Delta^\dagger$ is particularly easy in this case. In fact, by means of $\gamma^{\dagger\mu}=\gamma^0\gamma^\mu\gamma^0$ and $(\gamma^0)^2=1_{4\times4}$, it is straightforward to see that  
\begin{equation}
\Delta^\dagger\gamma^0=m^{-1}\gamma^0\gamma^\mu p_\mu,\label{3}
\end{equation} which is also the case for $\gamma^0\Delta$, so that the algebraic constraint is indeed respected for the parity case. The covariance is also respected for this case, as it can be readily seen from the fact that $\Delta=m^{-1}\gamma^\mu p_\mu=\Delta'$, therefore $[\Delta,S]=0$. Incidentally one could directly compute the commutator making use of the explicit form of the (infinitesimal) orthochronous proper, $L_+^\uparrow$, transformation $S\approx 1-\frac{i}{4}\delta\omega^{\alpha\beta}\sigma_{\alpha\beta}$, where $\delta\omega^{\alpha\beta}$ stands for the infinitesimal parameters and $\sigma_{\alpha\beta}$ denotes the transformation generator: 
\begin{equation}
[\Delta,S]=-\frac{i}{4}\delta\omega^{\alpha\beta}[m^{-1}\gamma^\mu p_\mu,\sigma_{\alpha\beta}]=-\frac{m^{-1}}{2}\Big(\delta\omega^{\alpha\beta}p_\beta\gamma_\alpha-\delta\omega^{\alpha\beta}p_\alpha\gamma_\beta\Big),\label{citada}
\end{equation} which amounts out to zero due to the antisymmetry of $\delta\omega^{\alpha\beta}$. Therefore a spinorial dual defined as $\Psi^\star=(P\Psi)^\dagger\gamma^0$ is consistent algebraically and physically. This result, along with the fact that $P \in \End(P_{SL(2,\mathbb{C})}\times_\rho\mathbb{C}^4)$, says that when $P\Psi=\tilde{\Psi}$ with $\tilde{\Psi}$ different from (and not proportional to) $\Psi$, but also a section of $P_{SL(2,\mathbb{C})}\times_\rho\mathbb{C}^4$, it is opened the possibility of composing the dual of a given spinor $\Psi$ with another spinor $\tilde{\Psi}$. This is not particularly new, but we would like to return to this point calling attention to an important consequence later on.

The fact that the dual composed with the parity operator trivially satisfies the algebraic and covariance constraints may give the impression of a not so restrictive scenario. This is not the case, however. Let's move on to the study of $\Delta=C$ and gather new impressions. The charge conjugation operator, acting upon spinors, may be written as $C=\gamma_2 K$, where $K$ is a complex conjugation operator acting from the left, so that $C^\dagger=\gamma_2^TK$. For this case, $\gamma^0\Delta=\gamma^0\gamma_2K$. Nevertheless, as $\gamma_2^T=\gamma_2$ and $\{\gamma^0,\gamma_2\}=0$, $\Delta^\dagger\gamma^0=-\gamma^0\Delta$ and the algebraic constraint is not satisfied. Moreover, in order to fill the covariance constraint one is lead to compare $S\gamma_2$ with $\gamma_2 S^*$ which, by its turn, leads to (the need for) an equality between $-\delta\omega^{\alpha\beta}\sigma_{\alpha\beta}\gamma_2$ and $\delta\omega^{\alpha\beta}\gamma_2\sigma^*_{\alpha\beta}$, something not reachable. Therefore $\Delta=C$ fails to accomplish both constraints. There is, however, a possible particular way to circumvent the constraints, which is the appreciation of eigenspinors of $C$. In fact, taking into account the action of $K$ the real eigenvalues of $C$ are simply the roots of the polynomial $(x^2-1)^2$, that is $\pm 1$. Hence a dual taking $C$ in its definition would lead to $\Psi^\star=(C\Psi)^\dagger\gamma^0=\pm(\Psi)^\dagger\gamma^0$, for which both constraints are trivially satisfied. This way out, however, may not be fully satisfactory from the physical point of view (see \cite{Ahl} for a set of eigenspinors of $C$, the elkos, whose standard dual leads to a null norm), but it may be suitable for Majonara spinors for instance.    

The situation of $\Delta=T=i\gamma_5\gamma_2K$ is similar\footnote{See \cite{Ahl} for a discussion about the form of $T$ acting upon spinors.} to the previous one. In fact, $\Delta^\dagger\gamma^0=i\gamma_2\gamma_5\gamma^0K$ and, employing the usual gamma anticommutators, we are lead to $\Delta^\dagger\gamma^0=-\gamma^0\Delta$ and the algebraic constraint is not satisfied. Analogously to the $C$ case, it is possible to show that the covariance is also lost. The main difference between this one and the previous case rests upon the fact that here the eigenspinor relation is of little help. The eigenvalues of $\Delta=T$ are given by $xi$, where $x$ is again given by the roots of $(x^2-1)^2$, that is $\pm i$. This imaginary unity, however, turns the algebraic constraint unreachable.  

The composition of discrete symmetries may also be studied following the above reasoning. Here we shall just tabulate the results with a broad brush. When $\Delta=CP$ the algebraic constraint is satisfied, while the covariance is not assured; $\Delta=PT$ does not even fill the algebraic constraint. The $\Delta=CT$ case performs an interesting situation to which we shall pay more attention. In fact, it is straightforward to see that in this case, the operator reduces to $\Delta=i\gamma_5$, for which the algebraic constraint is indeed readily satisfied. The covariance condition amounts out to give $[\Delta,S]=\frac{1}{4}[\gamma_5,\sigma_{\alpha\beta}]\delta\omega^{\alpha\beta}=0$ rendering a covariant dual. Nevertheless, since $\Delta=i\gamma_5$, this last accomplishment is only fully valid for $S\in L_+^\uparrow$, something which by the very nature of this paper we cannot claim. Thus, for Lorentz transformation not belonging to the orthochronous proper Lorentz subgroup, the transformation of $\Psi^\star$ would change its sign, rendering a pseudo-dual so to speak. Bilinear covariants would interchange positions: pseudo scalar and current would transform without sign and usual scalar, vector and bivector would be the pseudo quantities. Similar arguments and results used for $\Delta=CT$ apply for $\Delta=CPT$ and, of course, the results for $\Delta$ operators composed by any powers of these operators may be obtained from the analysis just outlined.

\section{Mapping spinors classes according to spinor duals}\label{IV}

Bearing in mind the previous discussions, in this section we look towards settling a relation (roughly speaking, a mapping program) among duals and spinor field classification. Our focus is to understand extensions of the usual Lounesto classification \cite{lou}, for which $\Delta = \mathbbm{1}$, and investigate classifications composed with unusual dual structures, in the light of \cite{beyondlounesto}, when $\Delta\neq \mathbbm{1}$. Such a task is accomplished by invoking duals built upon discrete symmetries and also their combinations. As far as we know, the evasion procedure of the dual structure advocated by Lounesto, is often a valuable and necessary  mathematical mechanism to unveil new and relevant physics associated with spinors, especially regarding the cases reported in \cite{Ahl, eplahl, dharamdipole}.

From a mathematical point of view, there is an important generalization of the inversion theorem \cite{taka}, recovering spinors from bilinear covariants, performed by the introduction of the so-called Fierz aggregate \cite{crawford1, crawford2, lou} $Z=\sigma+\mathbf{J}+i\mathbf{S}+\mathbf{K}\gamma_{5}-i\omega\gamma_{5}$, being $\sigma$, $\omega$, $\mathbf{J}$, $\mathbf{S}$ and $\mathbf{K}$ all the bilinear quantities associated to $\Psi$. The Fierz-Pauli-Kofink equations may be replaced by more restrictive quadratic equations involving $Z$, given by \cite{lou}  

\begin{eqnarray}\label{multi}
&&Z^{2}=4\sigma Z, \;\;\; Z\gamma_{\mu}Z=4J_{\mu}Z,\;\;\; Zi\gamma_{5}Z=4\omega Z,\\
&&Zi\gamma_{\mu}\gamma_{\nu}Z=4S_{\mu\nu}Z,\;\;\; Z\gamma_{5}\gamma_{\mu}Z=4K_{\mu}Z\nonumber.
\end{eqnarray}

Usually the definition of the Dirac dual structure is accomplished by the constraint $\Delta = \mathbb{1}$ in the general adjoint form introduced above. Nonetheless, guided by the discussions around Dirac spinors in Ref. \cite{rodolforegular}, more specifically for spinors belonging to class-2, which stand for eigenspinors of parity operator, the correct way to define the dual structure is performed via $\Delta \Psi = P\Psi = \pm\Psi$ --- recalling, once again, the very definition of the parity operator $P = m^{-1}\gamma^{\mu}p_{\mu}$ \cite{loh}. Convinced by the foregoing arguments, the correct Dirac dual structure should be interpreted as a structure in which the $P$ operator plays a central role. Thus, a quick inspection shows that when the parity operator is introduced in the dual structure it forces any spinor to belong to class-2 within its specific spinor classification \cite{beyondlounesto}. Such a result is general, encompassing any spinor and proving that the parity operator is closely connected with class-2 spinors, the specific class responsible for encompassing expansion coefficients ensuring locality for the quantum fields. Remarkable examples of what was previously discussed can be found in \cite{Ahl, rodolforegular, dharamdipole, eplahl, spinhalf}, showing the consistency of the comments around equation \eqref{citada}.   
 
Interestingly enough, for all $\Delta$ presented in this paper, the identities (\ref{multi}) are satisfied, enabling a classification for each dual case. We shall just report the main results here starting from the physical information encoded in the charge-conjugation operator. When setting $\Delta = \mathcal{C}$ on dual structure, leading, then, to $\Psi^\star=(C \Psi)^\dagger \gamma^0$, we are automatically taken to class-5. Finally, we investigate the dual structure defined in terms of the time-reversal operator, $\Psi^\star=(T \Psi)^\dagger \gamma^0$. For this particular case, the bilinear $J$ is always null, making it possible to access some extras classes, more specifically class-7 \cite{beyondlounesto, jotanulo}. Nonetheless, the physical information associated with such a class is still somewhat unknown. If we follow the previous line of reasoning, we expect that at least a subgroup of spinors belonging to this class should be somehow related to the $T$ operator. So far, the current literature has not presented any physical candidate to belong to such a class. 

For the interesting case of pseudo-dual investigated last section, $\Delta=CT$, all Lounesto classes are reachable, depending on the values of trial phases inserted in the spinor entries \cite{rodolforegular}. This last behavior is also accomplished by $\Delta=PT,PC,CPT$, with the additional remark that for $\Delta=PT$ class-7 can be also obtained.   
 
\section{Concluding Remarks: impacts and extensions}\label{remarks}

In the course of our analysis, we called attention to the fact that $\Delta\in \End(P_{SL(2,\mathbb{C})}\times_\rho \mathbb{C}^4)$ and, then, the unusual duals may be such that $\Psi^\star=\tilde{\Psi}^\dagger\gamma^0$, with $\tilde{\Psi}$ different from $\Psi$. We would like to outline a possible outcome from this freedom coming from the realm of Clifford algebra. For the discussion we are about to do, the results do depend on the spinor set at hand and not only on the operators acting upon them. Suppose then a given subset $\Omega$ of sections of $P_{SL(2,\mathbb{C})}\times_\rho \mathbb{C}^4$, or equivalently $\Omega \subset \mathbb{C}\otimes Cl_{1,3}F$, and let $\Delta \in\End(\Omega)$. Also, let $\Sigma_k(\Psi\Psi^\star)$ denotes the spin sums performed by the set of spinors belonging to $\Omega$ and its corresponding (but unusual) duals. The label $k$ stands for all the different sums necessary to the theory. The first aspect to be noted is that invariance of the norm does not ensure spin sum invariance or covariance in general. In fact, the transformed spin sums reads $\Sigma_k'(\Psi'\Psi'^\star)=S\Sigma_k(\Psi\Psi^\star)S^{-1}$ and a sufficient condition for invariance of $\Sigma_k$ is $[\Sigma_k(\Psi\Psi^\star),S]=0$ for each $k$. The usual exception is the well known case of Dirac spinors. In this last case, the spin sums are given by the Dirac operator. From the perspective of this paper, since the Dirac operator is identified with parity, the invariance of Dirac spinors spin sums may also be framed in the discussion around Eq. (\ref{citada}), which essentially asserts $[P,S]=0$. For any other case, the investigation of $[\Sigma_k(\Psi\Psi^\star),S]$ shall also be taken into account in the physical formulation, but we call attention to the case presented in Ref. \cite{eplahl}, for which the invariance of the spin sums are also respected. There is another remarkable point presented in \cite{eplahl}. We shall sketch this point here, trying to emphasize the formulation we have approached in the paper.        

Suppose the use of $\Omega\subset P_{SL(2,\mathbb{C})}\times_\rho \mathbb{C}^4$ elements as expansion coefficients of a given fermionic quantum field $\Phi[\Psi]$ and its corresponding adjoint $\Phi^\star[\Psi^\star=(\Delta \Psi)^\dagger\gamma^0]$ acting in a given Hilbert space. To work out quantum correlators is a demand of any field theory, a fact that underlies the relevance of spin sums and, thus, of the spinor dual. To fix ideas, let us particularize a bit this general analysis by fixing a set of four spinors in $\Omega$ along with, as usual, two spin sums. The space-like quantum correlator between $\Phi(\Psi)_{x}$ and $\Phi^\star(\Psi^\star)_{x'}$ $(x-x')<0$ is given by \cite{wei}
\begin{equation}
[\Phi(\Psi)_{x},\Phi^\star(\Psi^\star)_{x'}]\Big|_{(x-x')<0;(\mp)}=\Sigma_1\mathcal{H}(x'-x)\mp \Sigma_2 \mathcal{H}(x-x'),\label{i1}
\end{equation} where the up sign stands for the commutator, while the down sign denotes the anticommutator, and $\mathcal{H}(y)$ is the Hankel function. The sing in (\ref{i1}) must be chosen such that it vanish, in order to keep Lorentz invariance of the scattering matrix built with $\Phi(\Psi)$ and $\Phi^\star(\Psi^\star)$. Just recalling the standard case, for the Dirac field the spin sums are such that
\begin{eqnarray}
\Sigma_1\mathcal{H}(x'-x)|_{(x-x')<0}=-\gamma\cdot\partial\mathcal{H}(x'-x)-m\mathcal{H}(x'-x),\nonumber\\ \Sigma_2\mathcal{H}(x'-x)|_{(x-x')<0}=-\gamma\cdot\partial\mathcal{H}(x-x')+m\mathcal{H}(x-x') \label{i2}
\end{eqnarray} and, as the Hankel function is even (thus with first derivative odd), Eq. (\ref{i1}) amounts out to $[\Phi(\Psi)_{x},\Phi^\star(\Psi^\star)_{x'}]|_{(x-x')<0;(\mp)}=[-\gamma\cdot\partial+m\mp(\gamma\cdot\partial-m)]\mathcal{H}(x'-x)$. Therefore the plus sign must be taken, rendering the well known anti commuting statistics for the Dirac field. Of course, as stated, the statistics is utterly related to the spin sums at hand and, then, to the spinorial duals freedom. This observation is crucial to the appreciation of the spinor field possibility (with dual built with $\Delta=P$) investigated in Ref. \cite{eplahl}. In this last case
%In the manuscript, the spin is 1/2 (not \textcolor{red}{to be}  confused with fermions or bosons). The commutators in equation (12) lead to bosonic statistics, while anticommutators  yield fermionic statistics (still preserving the spin). This correction propagates to the end of the manuscript. So phrases like "commuting statistics for the fermionic  field" are incorrect. What the authors have in mind is something \textcolor{red}{like this}: we obtain commutators for a spin half field, that is a bosonic field. This is particularly important after equations (13) and (15).
\begin{eqnarray}
\Sigma_1\mathcal{H}(x'-x)|_{(x-x')<0}=A\cdot\partial\mathcal{H}(x'-x)+m\mathcal{H}(x'-x),\nonumber\\ \Sigma_2\mathcal{H}(x'-x)|_{(x-x')<0}=-A\cdot\partial\mathcal{H}(x-x')+m\mathcal{H}(x-x'), \label{i3}
\end{eqnarray} where $A=\phi\circ\gamma$ ($\phi$ being an algebraic homomorphism, not relevant here). The analogous of Eq. (\ref{i1}) for this case may be recast as
\begin{equation}
[\Phi(\Psi)_{x},\Phi^\star(\Psi^\star)_{x'}]\Big|_{(x-x')<0;(\mp)}=[A\cdot\partial+m\mp(A\cdot\partial+m)]\mathcal{H}(x'-x).\label{i4}
\end{equation} 
Bearing in mind the spin-$1/2$ framework, the minus sign must be taken forcing one to conclude in favour of a commuting statistics engendering another type of spin-$1/2$ field, namely spin-$1/2$ bosonic field. This fact was stressed in \cite{eplahl} and we are enforcing it to call attention to the relevance of the dual spinor theory in general as well as the use of discrete operators within $\Psi^\star$. In Ref. \cite{ij}, this kind of behaviour is also evinced, along with an interesting cosmological application.  

We would like to finalize this paper calling attention to some potentially interesting extensions of the analysis here performed. Firstly, as pointed out in the text, the identification of $\eta=\gamma^0$ is made when $\Delta$ is the identity operator\footnote{In Ref. \cite{Ahl} it is shown al least one more $\Delta$ operator, for the Elko dual case, different from the identity for which $\eta$ is also identified with $\gamma^0$.}. This identification is suitable since it allows for the usage of standard well-known relations involving $S(\Lambda)$. Without fixing it, it may well be the case of too many variables to handle and at least part of the analysis may be spoiled. Even so, a broader investigation may be undertaken. Finally, we call attention to a possibility which certainly deserves a further investigation: the creation of a dual space, $F(\mathbb{C}\otimes CL_{1,3})\supset \Omega^\star$ composed by elements as $\Psi_i^\star=a_i\Psi_j^\dagger\gamma^0$ with $i\neq j$ (exploring the consequences of $\Delta\in\End(\mathbb{C}\otimes CL_{1,3}F)$) with complex $\{a_i\}$, chosen in order to ensure a positive norm, orthogonality conditions and so on. This procedure needs a proper specific separation of the spinor space, with careful construction, but it would have the bonus of circumventing partially the constraints in dual spinorial construction. We shall delve into this possibility in the future.
\section*{Acknowledgments}	
	
	JMHS thanks to CNPq (grant No. 303561/2018-1) for financial support.

\end{document}